\documentclass[a4paper]{article}

\usepackage{INTERSPEECH2020}
\usepackage{xcolor}
\usepackage{arydshln}
\usepackage{subfiles}
\usepackage{hyperref}

\usepackage{tikz}
\usetikzlibrary{shapes.geometric, arrows}
\tikzstyle{pid} = [rectangle, rounded corners, minimum width=2cm, minimum height=1cm,text centered, text width=2cm, draw=black,  fill=red!30]
\tikzstyle{sid} = [rectangle, rounded corners, minimum width=2cm, minimum height=1cm,text centered, text width=2cm, draw=black, fill=green!30]
\tikzstyle{encoder} = [rectangle, rounded corners, minimum width=4cm, minimum height=1.5cm,text centered, draw=black, fill=blue!30]
\tikzstyle{arrow} = [ draw=blue, fill=blue!30,
                minimum height=7mm,shape border rotate=90]

\usepackage{setspace}
\twocolumn
\newcommand{\etal}{\textit{et al}.}
\title{Exploring the Use of an Unsupervised Autoregressive Model as a Shared Encoder for Text-Dependent Speaker Verification}

\name{Vijay Ravi, Ruchao Fan, Amber Afshan, Huanhua Lu, Abeer Alwan \thanks{\hspace{0.35em} This study was supported in part by the NSF.}}
\address{University of California Los Angeles, USA}
\email{(vijaysumaravi,fanruchao,amberafshan,huanhua,alwan)@ucla.edu}

\begin{document}

\maketitle
\begin{abstract}
In this paper, we propose a novel way of addressing text-dependent automatic speaker verification (TD-ASV) by using a shared-encoder with task-specific decoders. An autoregressive predictive coding (APC) encoder is pre-trained in an unsupervised manner using both out-of-domain (LibriSpeech, VoxCeleb) and in-domain (DeepMine) unlabeled datasets to learn generic, high-level feature representation that encapsulates speaker and phonetic content. Two task-specific decoders were trained using labeled datasets to classify speakers (SID) and phrases (PID). Speaker embeddings extracted from the SID decoder were scored using a PLDA. SID and PID systems were fused at the score level. There is a 51.9\% relative improvement in minDCF for our system compared to the fully supervised x-vector baseline on the cross-lingual DeepMine dataset. However, the i-vector/HMM method outperformed the proposed APC encoder-decoder system. A fusion of the x-vector/PLDA baseline and the SID/PLDA scores prior to PID fusion further improved performance by 15\% indicating complementarity of the proposed approach to the x-vector system. We show that the proposed approach can leverage from large, unlabeled, data-rich domains, and learn speech patterns independent of downstream tasks. Such a system can provide competitive performance in domain-mismatched scenarios where test data is from data-scarce domains.

\end{abstract}

%%%%%%%%%%%%%%%%%%%%%%%%%%%%%%%%%%%%%%
\section{Introduction}
%%%%%%%%%%%%%%%%%%%%%%%%%%%%%%%%%%%%%%
% \ruchao{How about 'Exploring the Use of Shared Speech Representation for...'}
% Text Dependent Speaker Verification, SDSVC 2020. 
% E2E, Encoder-Decoder, Triplet Loss for SV. 
% Unsupervised Training
% Explain the need for representation learning. 
% Novelty in this paper - APC for PID and SID

Text-dependent automatic speaker verification (TD-ASV) systems classify pairs of speech utterances as same or different based on the speaker's identity and the lexical content of the phrases spoken. This is analogous to two-factor authentication, in that the phrase identification (PID) and the speaker identification (SID), both, have to match for the user to gain access. The applications of TD-ASV include, but are not limited to, bio-metric verification in healthcare~\cite{sigona2018voice}, banking, forensics~\cite{singh2012applications}, and  privacy protection in personalized voice-assistants~\cite{changmy}. 

While the same-or-different speaker decision accuracy is of utmost importance, it is also beneficial if the TD-ASV system is resilient to domain mismatch between the training and testing data. This would enable the deployment of TD-ASV systems, originally developed for data-rich domains, to data-scarce domains thereby extending TD-ASV to unconventional domains like children's speech or zero-resource languages. To facilitate research in this direction, the short-duration speaker verification challenge (SDSVC), 2020~\cite{sdsvc2020plan} provides a standardized evaluation platform for researchers to test and benchmark their ASV systems using a common evaluation dataset. In this study, we address the problem of TD-ASV in a novel way by training an encoder in an unsupervised fashion to learn shared feature representations of both speaker and phrase identity.   

\begin{figure}[h]
    \centering
    \includegraphics[width=\linewidth]{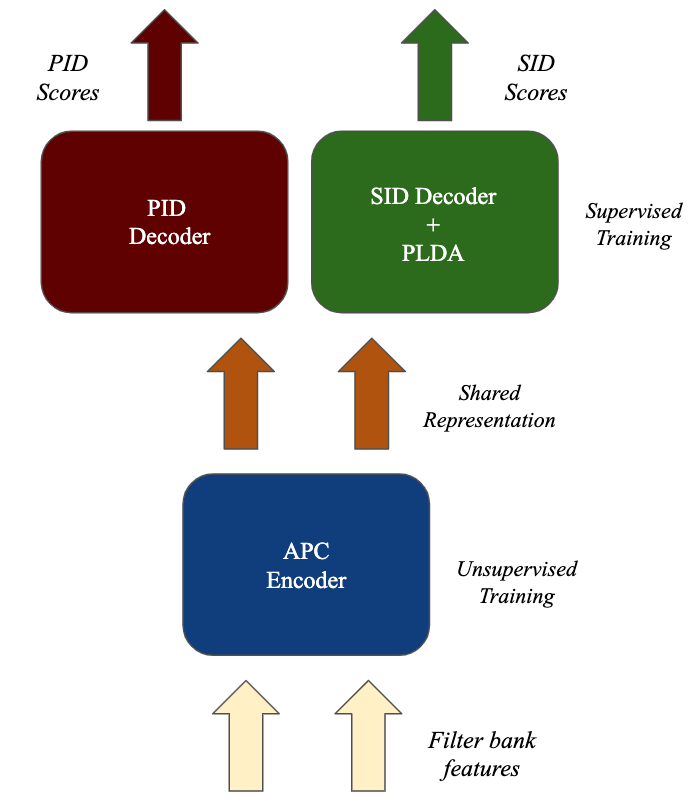}
    \caption{\label{fig:model_arch} Encoder-Decoder model architecture proposed in this paper. The encoder is an APC model trained in an unsupervised way to learn a generic, high-level feature representation independent of downstream tasks. The decoders (PID and SID) are trained in a supervised manner.}
\end{figure}

Previously, the i-vector/PLDA (probabilistic linear discriminant analysis) method~\cite{dehak2010front,kenny2013plda} and some of its extensions~\cite{larcher2013phonetically,stafylakis2013text} showed promising results on the TD-ASV task. Zenali~\etal~introduced the HMM based i-vector approach~\cite{zeinali2016vector,zeinali2017hmm}, and used a set of phone-specific HMMs to collect the statistics for i-vector extraction. In~\cite{variani2014deep}, Variani~\etal~replaced the conventional i-vectors by using deep neural networks (DNNs) to learn speaker discriminative features (d-vector). A phonetically-aware TD-ASV system was developed to extract i-vectors using: a) output posteriors~\cite{lei2014novel} and b) bottleneck features~\cite{zeinali2016deep}, as frame alignments, which were generated from a DNN trained for automatic speech recognition (ASR).To tackle the shorter utterance problem, convolutional neural networks~\cite{guo2017cnn} and DNNs~\cite{guo2018deep} were used to map the i-vectors extracted from
short utterances to the corresponding long-utterance i-vectors. Although these systems were effective, they relied on handcrafted dictionaries to generate alignments for every phrase and large, labeled, in-domain datasets. On the contrary, the proposed method needs no dictionaries or alignments and can take advantage of abundantly available out-of-domain data. More recently, the end-to-end (E2E) approach of training TD-ASV systems has gained significant momentum. Heigold~\etal~proposed an E2E system combining the training, the evaluation and the verification process into a single compact network and jointly optimized all parameters using a verification-based loss~\cite{heigold2016end}. In~\cite{zhang2016end}, Zhang~\etal~suggested an attention based E2E network for jointly learning speaker and phonetic discriminative features. In contrast to the previous E2E systems that were trained on a tuple-based loss function, Wan~\etal~proposed the generalized E2E loss function~\cite{wan2018generalized}. These E2E systems were, however, computationally expensive and optimized to perform well only for a specific phrase (eg: the wake-word phrase).

%To tackle the shorter utterance problem, convolutional neural networks~\cite{guo2017cnn} and DNNs~\cite{guo2018deep} were used to learn a mapping from short and long utterances.
 
% In addition, the E2E systems have not been evaluated on non-homogeneous, cross-lingual datasets. 

Inspired by the recent success of unsupervised pre-training~\cite{devlin2018bert, schneider2019wav2vec} and representation learning~\cite{chung2019unsupervised,chung2020generative,oord2018representation}, we propose to use a shared-encoder with two task-specific decoders for TD-ASV. The model architecture is as shown in Figure~\ref{fig:model_arch}. %use ~ with \ref or \cite
Specifically, an autoregressive  predictive coding (APC) encoder~\cite{chung2019unsupervised} is trained in an unsupervised way to learn a generic feature representation. The encoded representation encapsulates both speaker and phonetic discriminative features. We then use features extracted from the encoder as input to task-specific decoders to predict phrase identity and extract speaker embeddings. Since the APC encoder is trained using unlabeled data (in-domain and out-of-domain), it is capable of capturing high-level speech representation independent of data domain or downstream tasks. The proposed shared-encoder architecture obviates the need for two separate encoders for each individual task and large amounts of labeled in-domain data for training. Results on the domain-mismatched evaluation data demonstrate that the proposed shared-encoder model can also be effective in domain adaptation in TD-ASV.  %\vijay{rewrite hypothesis to be a little more general so it includes domain adaptation.}  

Prior work in  feature-learning includes~\cite{liu2015deep, chung2019unsupervised,sarkar2019time}. Liu~\etal~suggested the use of DNNs for feature extraction~\cite{liu2015deep}. Their method, however, required labeled data for training the feature-extractor in contrast to the unsupervised method employed in this study. Chung~\etal~proposed the unsupervised APC encoder in ~\cite{chung2019unsupervised} but used the extracted feature representations with an i-vector/PLDA SID system as opposed to the task-specific decoders suggested in this paper. While these methods reported results on domain-matched datasets, the proposed model was evaluated on DeepMine data which consists of Persian and English phrases spoken by non-native English speakers. All evaluations are in accordance with Task-1 of the short duration speaker verification challenge (SDSVC)~\cite{sdsvc2020plan}. 

The remainder of the paper is organized as follows: in Section~\ref{sec:encoder_decoder}, the encoder-decoder structure is presented. The datasets used and the model architecture proposed are outlined in Section~\ref{sec:exp}, results are presented and discussed in Section~\ref{sec:results} and conclusion and the future directions are provided in Section~\ref{sec:conclustion}.

% \subsection{Previous work}
%%%%%%%%%%%%%%%%%%%%%%%%%%%%%%%%%%%%%%%%%%%%%%%%%%%%%%%%%%%%%%%%%%
\section{\label{sec:encoder_decoder}Encoder-Decoder TD-ASV}
%%%%%%%%%%%%%%%%%%%%%%%%%%%%%%%%%%%%%%%%%%%%%%%%%%%%%%%%%%%%%%%%%%
%%%%%%%%%%%%%%%%%%%%%%%%%%%%%%%%%%%%%%%%%%%%%%%%%%%%%%%%%%%%%%%%%%
\subsection{\label{ssec:APC}Autoregressive Predicitive Coding (APC) Encoder}
%%%%%%%%%%%%%%%%%%%%%%%%%%%%%%%%%%%%%%%%%%%%%%%%%%%%%%%%%%%%%%%%%%
Predictive coding has played an important role in speech processing, especially in speech coding using linear prediction coding (LPC)~\cite{o1988linear}. LPC predicts future audio samples whereas, a recently proposed autoregressive predictive coding~\cite{chung2019unsupervised} predicts the features of a future frame. The idea is to utilize the input sequence itself as labels and predict a frame $n$ steps ahead of the current frame to achieve unsupervised speech representation learning. The model architecture is as shown in Figure~\ref{fig:model_arch}.  %We therefore train the encoder with the APC mechanism to extract shared speech representation for both phrase-id and speaker-id tasks. 

% \begin{figure}[h]
%     \centering
%     \includegraphics[width=\linewidth]{LaTeX/figures/model_architecture.png}
%     \caption{\label{fig:model_arch} Encoder-Decoder model architecture proposed in this paper. The encoder is an APC model trained in an unsupervised way to learn a generic, high-level feature representation independent of downstream tasks. The decoders (PID and SID) are trained in a supervised manner.}
% \end{figure}

Suppose the input speech sequence is $\textbf{X}=(x_1, x_2,...,x_T)$, the time shift of prediction is fixed at $n$, and the ground truth of the prediction for each frame is $(x_{1+n}, x_{2+n},...,x_{T+n})$. In order to prevent the model from learning a trivial solution, we apply a uni-directional neural network structure, as opposed to bi-directional networks, by letting the model be aware of the context only from history. By stacking multiple long short-term memory (LSTM) layers and adding residual connections, we obtain a deep LSTM network. Prior to that, a two-layer feed-forward network is considered as the pre-net network to transform the speech features into a hidden latent space. Together with LSTMs, we denote this combined network as DLSTM. The output of the DLSTM is then fed into a linear layer and transferred to the input space, which means that the dimension will be the same as the input features. Mathematically, the model architecture can be described as follows:

\begin{equation}
  \textbf{Y} = W_f DLSTM(\textbf{X}, W_{lstm}) + b_f
  \label{eq:encoder}
\end{equation}

where $W_{lstm}$ represents all the parameters in the DLSTM; $W_f$ and $b_f$ denote the weight matrix and bias vector in the last layer, respectively; and $\textbf{Y}=(y_1, y_2,...,y_T)$ is the output. Considering the L1 loss as a metric distance for prediction, all the above parameters are obtained by optimizing the following loss function:
\begin{equation}
  L_{1} = \sum_{t=1}^{T-n}|x_{t+n}-y_t|
  \label{eq:l1loss}
\end{equation}

% After training the final model, the last mapping layer is removed and the weights of the DLSTM are used as the speech representation of each utterance. 

% \subsection{Phrase-ID Decoder}
\subsection{Task-specific Decoder}
\label{ssec:ts-decoder}
The PID decoder was designed to distinguish between different phrases. In order to obtain better generalization and faster convergence, we allowed the PID decoder to learn frame-level phonetic representations through a phoneme classification task using the connectionist temporal classification (CTC)~\cite{graves2006connectionist}. The frame-level representations were then averaged using a statistical pooling layer to form a single feature vector for sentence-level phrase classification. Specifically, the speech representation obtained in Section~\ref{ssec:APC} was first fed into a stacked bi-directional LSTM network (BLSTM) to get the frame-level representations. Then, the frame-level representations were used as the inputs for two subsequent networks. In the first network, they were transformed into the phoneme space to capture phonetic information. In the second network, a pooling layer and two feed-forward layers were used to transcribe the frame-level representations to phrase-ID space followed by a softmax layer. The overall PID decoder was optimized by jointly minimizing the following loss:

\begin{equation}
    \mathcal{L}_{total} = \mathcal{L}_{CTC} + \lambda \mathcal{L}_{CE},
\end{equation}

where, $\mathcal{L}_{CTC}$ is the CTC loss for phoneme classification and $\mathcal{L}_{CE}$ is the loss arising from the phrase classification. We use $\lambda$ as a regularizing hyperparameter to control the contribution of the CE loss to the total loss. 

% Another finding is that  $\lambda$ does not have to be large to achieve a good performance for phrase classification, which means that the CTC is the main task. \Amber{So? why do you use CE then or what do you conclude by saying CTC is the main loss. Maybe this lambda part doesn't belong here and in the discussion. Decide depending on the flow. It will be ill fit there but talking about performance in section 2 umm... }

% \subsection{Speaker-ID Decoder}
The speaker-ID decoder consists of another BLSTM network followed by a statistical pooling layer to extract speaker embeddings. Speech representations obtained from the APC encoder in Section~\ref{ssec:APC} are used here as input. The size of the final transformation layer is dependent on the number of speakers in the dataset. The SID decoder is optimized by minimizing the cross entropy loss arising from the classification of speakers. 
%%%%%%%%%%%%%%%%%%%%%%%%%%%%%%%%%%%%%%%
\section{\label{sec:exp}Experimental Details}
%%%%%%%%%%%%%%%%%%%%%%%%%%%%%%%%%%%%%%%
\subsection{Datasets}
%%%%%%%%%%%%%%%%%%%%%%%%%%%%%%%%%%%%%%%
The specifications of the datasets used in this paper are provided in Table~\ref{tab:data_shared_encoder}. Utterances from LibriSpeech, VoxCeleb1 and VoxCeleb2~\cite{Nagrani17} and DeepMine Part-1~\cite{zeinali2018deepmine,zeinali2019multi} were used for three different tasks: 1) Unsupervised pre-training of the shared encoder, 2) Phrase ID training, and 3) Speaker ID training. In this section, we provide details of the subsets of data used for each task.

%% Please add the following required packages to your document preamble:
% \usepackage{graphicx}
\begin{table}[h]
\centering
\caption{Details of the datasets used.}
\label{tab:data_shared_encoder}
\resizebox{\linewidth}{!}{%
\begin{tabular}{llrrr}
\hline \hline
\multicolumn{1}{c}{\textbf{Subset}} & \multicolumn{1}{c}{\textbf{Database}} & \multicolumn{1}{c}{\textbf{\begin{tabular}[c]{@{}c@{}}\#\\ Utts\end{tabular}}} & \multicolumn{1}{c}{\textbf{\begin{tabular}[c]{@{}c@{}}\#\\ Spks\end{tabular}}} & \textbf{\begin{tabular}[c]{@{}r@{}}Duration\\ (in hours)\end{tabular}} \\ \hline \hline \\
\textit{train-librispeech} & Librispeech & 140k & 5466 & 478.5 \\
\textit{dev-librispeech} & Librispeech & 2.7k & 97 & 5.3 \\
\textit{train-voxceleb} & VoxCeleb & 1.2M & 7350 & 2637.8 \\
\textit{dev-voxceleb} & VoxCeleb & 73k & 7350 & 151.2 \\
\textit{train-deepmine} & DeepMine & 101k & 963 & 91.5 \\
\textit{dev-deepmine} & DeepMine & 37k & NA & 31.6 \\ 
\textit{test-deepmine} & DeepMine & 69k & NA & 61.2 \\ \hline \hline
\end{tabular}%
}
\end{table}

The in-domain training data (\textit{train-deepmine}) contains speech utterances from 963 speakers, some of whom have only Persian phrases. The enrollment (\textit{dev-deepmine}) and test utterances (\textit{test-deepmine}) are drawn from a fixed set of ten phrases consisting of five Persian and five English phrases, respectively. More details of the phrases can be found in~\cite{zeinali2018deepmine}.

%%%%%%%%%%%%%%%%%%%%%%%%%%%%%%%%%%%%%%%%%%%%%%%%%%%%%%%%%%%%%%%%%%
\subsubsection{Unsupervised Pre-training of Shared Encoder}
%%%%%%%%%%%%%%%%%%%%%%%%%%%%%%%%%%%%%%%%%%%%%%%%%%%%%%%%%%%%%%%%%%
The unsupervised pre-training of the shared encoder used the out-of-domain \textit{train-librispeech} subset, 500k utterance from VoxCeleb and the in-domain \textit{train-depmine} subset. Since the APC encoder can be trained with unvoiced frames as well, no speech activity detection (SAD) is applied. A uniform sampling rate of 16~KHz is used across datasets. To prevent overfitting, a combined development set consisting of \textit{dev-librispeech}, \textit{dev-voxceleb} and \textit{dev-deepmine} were used for hyperparameter selection. 
%%%%%%%%%%%%%%%%%%%%%%%%%%%%%%%%%%%%%%%%%%%%%%%%%%%%%%%%%%%%%%%%%%
\subsubsection{Task Specific Decoder Training}
%%%%%%%%%%%%%%%%%%%%%%%%%%%%%%%%%%%%%%%%%%%%%%%%%%%%%%%%%%%%%%%%%%
For training the phrase ID decoder, 100 hours of LibriSpeech and all utterances of \textit{train-deepmine} were used. \textit{dev-librispeech} and the \textit{dev-deepmine} dataset were used for hyperparameter selection.

The SID decoder was trained using 1.2M utterances (7350 speakers) from the VoxCeleb dataset. Similar to the data processing of the x-vector system in~\cite{snyder2018x}, the utterances were cut into 3~second segments and augmented with noise from the MUSAN database~\cite{snyder2015musan} resulting in a total of 3.2M utterances ($\sim7$k hours). 
%The in-domain DeepMine data consisting of 101k utterances from 963 speakers were used to fine-tune the trained speaker ID decoder. For the unsupervised PLDA-adaptation, 37k utterances from the development set of the DeepMine data were used.

%%%%%%%%%%%%%%%%%%%%%%%%%%%%%%%%%%%%%%%%%%%%%%%%%%%%%%%%%%%%%%%%%%
\subsection{Front-End Processing}
%%%%%%%%%%%%%%%%%%%%%%%%%%%%%%%%%%%%%%%%%%%%%%%%%%%%%%%%%%%%%%%%%%
The Kaldi framework~\cite{povey2011kaldi} was used for all front-end preprocessing and feature extraction for each of the three tasks. The features are 40 dimensional filterbanks with a frame-length of 25ms and a frame shift of 10ms. Cepstral mean and variance normalization is applied on the features. The energy SAD (from Kaldi), used in the speaker embedding extraction, filters out non-speech frames. 
%%%%%%%%%%%%%%%%%%%%%%%%%%%%%%%%%%%%%%%%%%%%%%%%%%%%%%%%%%%%%%%%%%
\subsection{Model Architecture}
%%%%%%%%%%%%%%%%%%%%%%%%%%%%%%%%%%%%%%%%%%%%%%%%%%%%%%%%%%%%%%%%%%
\subsubsection{APC Encoder}
%%%%%%%%%%%%%%%%%%%%%%%%%%%%%%%%%%%%%%%%%%%%%%%%%%%%%%%%%%%%%%%%%%
% \ruchao{Make it pre-trained encoder}
The APC encoder DLSTM is composed of 4 layers of unidirectional LSTMs with each layer consisting of 512 hidden units. The input to the shared-encoder is 40 dimensional filter-bank features. The  shared encoder is trained in an auto-regressive manner by minimizing the L1 loss function as described in Section~\ref{sec:encoder_decoder}.

The pre-net feature embedding network of the encoder DLSTM is made up of 2 fully-connected layers with ReLU activations. The encoder model is initialized using the Xavier uniform initialization and a dropout of 0.1 is applied to the ReLu activation function. 

During evaluation, the shared-encoder is used as a feature extractor to extract learned representations for each utterance. These feature representations are the hidden RNN states of the APC model and form a 4-dimensional tensor of the shape (number-layers, batch-size, sequence-length, RNN-hidden-size). In our experiments, 512 dimensional hidden states of all 4 RNN layers of the APC model were used. Features extracted from the APC model are then fed into the task-specific decoder for learning the corresponding speaker and phrase identities.  

%%%%%%%%%%%%%%%%%%%%%%%%%%%%%%%%%%%%%%%%%%%%%%%%%%%%%%%%%%%%%%%%%%
\subsubsection{Task Specific Decoders}
%%%%%%%%%%%%%%%%%%%%%%%%%%%%%%%%%%%%%%%%%%%%%%%%%%%%%%%%%%%%%%%%%%
Two standalone decoders are trained to classify speech utterances based on speakers and phrase-IDs. Each decoder is trained and evaluated separately. 

The phrase ID (PID) decoder is composed of 3 layers of bidirectional LSTMs made up of 512 hidden units. The output of these BLSTM layers is then fed into two different sub-networks to predict phonemes and classify phrases. The mapping from Persian to English phoneme set is adopted as suggested in the data corpus, leading to 39 phonemes in total. Therefore, the phoneme prediction sub-network is a linear layer with a 40 dimensional (39 phonemes + 1 blank) output. The phrase classification sub-network consists of a pooling layer followed by a fully-connected layer (400 hidden units) and a prediction layer of 11 outputs (10 phrases + 1 no match). Since we utilize out of domain data which do not have phrase-ID labels, we add an extra category for all utterances whose contents do not match the given 10 phrases of the evaluation data. We observe that the PID decoder converges well when $\lambda$ (defined in section~\ref{ssec:ts-decoder}) is heuristically set to 0.2. %No further experiments are conducted for comparison in terms of $\lambda$, since the PID decoder converges well with this initial setting.

The speaker ID decoder is made up of 3 layers of bidirectional LSTMs each consisting of 512 hidden units. This is followed by statistical pooling, a fully-connected (dense) layer, and a prediction layer. The dimension of the prediction layer 7350 based on the number of speakers in the training set. During evaluation, the bottleneck features (outputs from the dense layer of the SID decoder) are extracted and used as speaker embeddings. The dimension of the fully-connected dense layer is set at 600 similar to the x-vector system. 

%%%%%%%%%%%%%%%%%%%%%%%%%%%%%%%%%%%%%%%%%%%%%%%%%%%%%%%%%%%%%%%%%%
\subsection{Model Training and Evaluation}
%%%%%%%%%%%%%%%%%%%%%%%%%%%%%%%%%%%%%%%%%%%%%%%%%%%%%%%%%%%%%%%%%%

The shared encoder was trained for 5 epochs with a learning rate of $2e^{-4}$. The weights and biases of the shared-encoder network were frozen after the training to ensure that the task-specific optimization of the decoders did not modify the shared-encoder the. Both the phrase ID and the speaker ID decoder networks were trained in parallel to minimize their corresponding loss functions. Decoders were trained for 5 epochs with a learning rate of $2e^{-4}$ and the learning rate was annealed by a factor of 0.5 after 3 epochs.   

% In the phrase ID decoder, the network predicted the ID of the spoken phrase during training, and during evaluation the phrase ID of the enrollment and test utterances were compared to predict scores for each pair. During the training phase, the decoder network is optimized to classify the speaker ID of the speech utterance. In the evaluation phase, we extract the speaker embeddings of the enrollment and test utterances. All model parameters were chosen heuristically. 

% In the Speaker ID decoder network, after training for 5 epochs, the decoder was fine-tuned with DeepMine data for another 5 epochs with a learning rate of $1e^{-4}$. While during training, the decoder network predicted the speaker ID of the speech utterance, for evaluation, we extract the embeddings of the enrollment and test utterances. \Amber{Don't you do this at development stage as well. I mean to say tell at some point what all the development set was used for}

% During the evaluation phase, the phrase ID of the enrollment and test utterances were compared to predict the PID scores for each pair.\ruchao{This expression is not clear. it should be "the log likelihood of each test utterance to be the phrase of the corresponding enrollment data is computed as the PID score." } 

During evaluation,the log likelihood of phrase-ID of test utterance and the corresponding enrollment utterance being the same is computed as the PID score. Speaker embeddings are extracted from the dense layer of the SID decoder. A PLDA classifier is used to compare the extracted speaker embeddings, and predict target/imposter speaker decisions. Speaker embeddings extracted from the speaker ID decoder were centered and projected using LDA. The LDA dimension was tuned on the VoxCeleb training set to 200. After dimensionality reduction, the representations were length-normalized and modeled by the PLDA and the PLDA model was then adapted using the DeepMine training data. The log-likelihood scores of the PLDA model (SID scores) and the PID model were fused to generate the final system prediction. 
% \Amber{I noticed in the SDSV doc and ivector/HMM paper it is target or impostor and for PID it is correct/wrong, so target-correct, target-wrong, imposter-correct and imposter-wrong. Maybe you want to stick to an accepted terminology.} 
%%%%%%%%%%%%%%%%%%%%%%%%%%%%%%%%%%%%%%%%%%%%%%%%%%%%%%%%%%%%%%%%%%
\section{\label{sec:results}Results and Discussion}
%%%%%%%%%%%%%%%%%%%%%%%%%%%%%%%%%%%%%%%%%%%%%%%%%%%%%%%%%%%%%%%%%%
Table~\ref{tab:results_sdsvc_task1} provides results obtained from the text-dependent speaker verification task of SDSVC on the evaluation data. System performance is compared using the normalized minimum detection cost function (minDCF)~\cite{martin2010nist}. %We also report the equal error rate (EER).  
% \subsection{Baseline Performance}

Two baselines were provided in the challenge evaluation plan for this task: the x-vector system and i-vector/HMM system. The state-of-the-art x-vector method, based on the TDNN architecture of~\cite{snyder2018x}, was trained using VoxCeleb1 and VoxCeleb2 databases.  Evaluation trials, as per the provided baseline, were scored using the PLDA without any score normalization. The i-vector/HMM method, that also takes into consideration phrase information, was selected as the second baseline. Among the published results, the i-vector/HMM method is the best performing system on DeepMine data.

% % Please add the following required packages to your document preamble:
% % \usepackage{graphicx}
% \begin{table}[h]
% \centering
% \caption{Results for text-dependent task of the SDSV Challenge. The best performing system is boldfaced. + indicates cascading of systems. \Amber{Fix this table}}
% \label{tab:results_sdsvc_task1}
% \resizebox{\linewidth}{!}{%
% \begin{tabular}{llcrr}
% \hline \hline
% \textbf{System} & \textbf{Classifier} & \textbf{Score Fusion} & \multicolumn{1}{c}{\textbf{minDCF}} & \multicolumn{1}{c}{\textbf{EER (\%)}} \\ \hline \hline \\
% i-vector/HMM$^*$ & PLDA & None & \textbf{0.1472} & \textbf{3.47} \\ \\
% x-vector$^*$ & PLDA & None & 0.5611 & 10.13 \\ \\ 
% % i-vector & PLDA & None & 0.5891 & 11.27 \\ \\
% x-vector & PLDA & PID & 0.2170 & 4.80    \\ \\
% % i-vector & PLDA & PID & 0.2589  & 6.03 \\ \\
% \begin{tabular}[c]{@{}l@{}}Shared Encoder \\ + SID Decoder \end{tabular} & \begin{tabular}[c]{@{}l@{}}PLDA  \\ + adaptation \\ with train\_deepmine\end{tabular} & PID & 0.2697 & 6.28 \\ \\
% \begin{tabular}[c]{@{}l@{}}Shared Encoder \\ + SID Decoder \\ + SID fine-tuning \end{tabular} & \begin{tabular}[c]{@{}l@{}}PLDA \\ + adaptation  \\with dev\_deepmine\end{tabular} & PID & 0.2464 & 6.61 \\ \hline \hline \\
% $^*$ indicates baseline. \\ + indicates cascading.

% \end{tabular}%
% }
% \end{table}

% \usepackage{graphicx}
\begin{table}[h]
\centering
\caption{Results for the text-dependent task of the SDSV challenge in terms of minDCF and EER. $^*$ indicates baseline and $+$ indicates score-level fusion using linear regression.}
\label{tab:results_sdsvc_task1}
\resizebox{0.9\linewidth}{!}{%
\begin{tabular}{lcrr}
\toprule
\toprule
\textbf{\begin{tabular}[c]{@{}c@{}}Speaker ID \\ System\end{tabular}} & \textbf{\begin{tabular}[c]{@{}c@{}}Phrase ID  \\ System \end{tabular}} & \multicolumn{1}{c}{\textbf{minDCF}} & \multicolumn{1}{c}{\textbf{\begin{tabular}[c]{@{}c@{}}EER\\  (\%)\end{tabular}}} \\ \midrule \midrule
x-vector$^*$ & None & 0.5611 & 10.13 \\ 
i-vector$^*$ & HMM & 0.1472 & 3.47 \\ \hdashline[2pt/2pt] 
x-vector & PID & 0.2170 & 4.80 \\ 
SID & PID & 0.2697 & 6.28 \\ \hdashline[2pt/2pt]
SID + x-vector & PID & 0.1830 & 4.18 \\ 
\bottomrule 
\bottomrule
\end{tabular}
}
\end{table}

The proposed system achieves a minDCF of 0.2697 and an EER of 6.28\%. This represents a relative improvement of 51.9\% in terms of minDCF (0.5611 for the x-vector baseline versus 0.2697 for the proposed method) and 38\% in terms of EER (10.13\% to 6.28\%). In order to have a fair comparison between the x-vector system and the shared-encoder system, we fused the scores of x-vectors and PID. We observed that, in this case, the performance of the fused x-vectors was better than the shared encoder system. The minDCF improved relatively by 19.5\% (from 0.2697 to 0.2170) and the EER by 23.5\% (from 6.28\% to 4.8\%). Thus, the x-vector system, on its own, is better at capturing speaker discriminatory features, than the SID network of the proposed framework. Nevertheless, on the overall task of TD-ASV, the proposed system performs better than the x-vector baseline. This improvement in performance can be attributed to the unsupervised pre-training of the shared-encoder using unlabeled in-domain data and the use of phonetic information by the proposed system. As a result, our system is better suited for the text-dependent, cross-lingual task of this challenge in comparison to the x-vector baseline.

%Strong evidence of the complementarity of our proposed approach was found when 
To further analyze the performance of the proposed system, fusion of the x-vector/PLDA scores and the SID/PLDA scores was performed using linear regression before fusing with PID scores. Equal coefficients of $0.5$ were chosen for this linear regression which resulted in a 15\% gain in minDCF (0.2170 to 0.1830) and a 12\% relative gain in EER (4.8\% to 4.18\%). These results seem to suggest that the SID system offers complimentary information to the x-vector system. It is possible that the proposed unsupervised method learns useful speaker-discriminative information that was previously discarded when learning representations in a supervised fashion. Combining supervised and unsupervised feature representations can therefore be advantageous in developing robust TD-ASV systems. 

% % Please add the following required packages to your document preamble:
% % \usepackage{graphicx}
% \begin{table}[h]
% \centering
% \caption{Results for TD-ASV in terms of minDCF and EER when the x-vector+PID system is fused with the SID+PID system. The numbers in the first two columns are the weights for fusion. The best performing configuration is boldfaced.}
% \label{tab:final_fusion}
% \resizebox{0.9\linewidth}{!}{%
% \begin{tabular}{cccc}
% \toprule
% \toprule
% \textbf{X-vector + PID} & \textbf{SID + PID} & \textbf{minDCF} & \textbf{EER} \\
% \midrule
% \midrule
% 1 & 0 & 0.2170 & 4.80 \\
% 0 & 1 & 0.2697 & 6.28 \\
% 0.8 & 0.2 & TBD & TBD \\
% 0.5 & 0.5 & 0.1830 & 4.18 \\
% 0.2 & 0.8 & TBD & TBD \\ \bottomrule \bottomrule
% \end{tabular}%
% }
% \end{table}

The performance of the i-vector/HMM method, on the other hand, exceeded that of the proposed method by 45\% (minDCF of 0.1472 vs 0.2697). This system used hidden Markov model (HMM) states to model time sequences and extract i-vectors for each phrase. The i-vector/HMM approach outperforms the proposed method mainly because of its capability to reject target-wrong trials, meaning that if two different phrases were spoken by the same speaker, the HMM Viterbi decoding produced invalid statistics for such trials and consequently they were rejected easily~\cite{zeinali2017hmm}. In contrast, since the PID and the SID systems were fused by a simple score-level fusion, our system may have predicted higher log-likelihoods. A comprehensive analysis of the results could not be performed because the ground truth labels for the evaluation data were not available.

% \Amber{Maybe write this more the differences in the approach and then the better performance. So the person reading this doesn't think we are contributing nothing.} The performance of i-vector/HMM method exceeds that of the proposed method by 39\% (minDCF of 0.1472 vs 0.2464). This system used hidden Markov model (HMM) states to model time sequences and extract i-vectors for each phrase. The i-vector/HMM approach outperforms the proposed method mainly because of its capability to reject target-mismatched trials, meaning that if two phrases are spoken by the same speaker, the Viterbi forced alignment produced invalid statistics for such trials and consequently they are rejected easily. \Amber{The following sentence might not be a good idea to start with in contrast.} In contrast, since the scores are fused by pair-wise dot product, our system may have predicted higher log-likelihoods. Further analysis of the results was not performed because the ground truth labels for the evaluation data were not available. 
% \Amber{I think there method is not robust to train test mismatch in text, and also channel stuff. I looked at the conclusion and some part of discussion in their papers. They are even getting improvement by fusing with ivector/GMM. Maybe you can see if you can contrast on one of these drawbacks, you can achieve better their.}

%\Amber{You have to repeatedly highlight that you are doing better than one of the baselines}
%%%%%%%%%%%%%%%%%%%%%%%%%%%%%%%%%%%%%%%%%%%%%%%%%%%%%%%%%%%%%%%%%%
\section{\label{sec:conclustion}Conclusion} 
%%%%%%%%%%%%%%%%%%%%%%%%%%%%%%%%%%%%%%%%%%%%%%%%%%%%%%%%%%%%%%%%%%
In this paper, a novel model architecture comprised of a shared-encoder with task-specific decoders was proposed for TD-ASV. An auto-regressive predictive coding encoder was trained in an unsupervised fashion to learn generic features independent of the downstream task. Task-specific decoders were then optimized for phrase and speaker classification. An improvement of 52\% was achieved in terms of minDCF compared to the x-vector baseline. The i-vector/HMM method was the best performing system. 

The proposed method has the advantage of learning high-level speech patterns from large, unlabeled, data-rich domains. The encoded speech representations successfully captured speaker and phonetic discriminative features. Results obtained on the evaluation dataset demonstrated the domain-adaptaion ability of the proposed system. Further, strong evidence of the complementarity of the proposed system was found when the x-vector scores were fused with the scores of the encoder-SID decoder.

% \Amber{Can we talk something about the accent problem in the dataset?}\vijay{I'll have to discuss that in introduction also. }

% Other domains that can benefit from the proposed approach include  
% % More research using the proposed approach on the various other databases, and preferably larger ones could provide insights regarding the generalization of this approach. 
A natural progression of this work is to compare the effectiveness of the APC encoder against other unsupervised methods such as the contrastive prediction approach. Further research could also be conducted to determine the applicability of the shared-encoder on other data-scarce domains, for example, accented speech, zero-resource languages, children's speech. Additionally, both PID and SID systems could be jointly trained as a multi-task problem to make the system more robust. %Results could be analyzed further to identify and resolve the limitations of the proposed framework.  

% %%%%%%%%%%%%%%%%%%%%%%%%%%%%%%%%%%%%%%%%%%%%%%%%%%%%%%%%%%%%%%%%%%
% \section{Acknowledgement}
% %%%%%%%%%%%%%%%%%%%%%%%%%%%%%%%%%%%%%%%%%%%%%%%%%%%%%%%%%%%%%%%%%%
% This study was supported in part by the NSF.
% \input{LaTeX/flowchart}
\bibliographystyle{IEEEtran}
\bibliography{mybib}

\begin{thebibliography}{10}
\providecommand{\url}[1]{#1}
\csname url@samestyle\endcsname
\providecommand{\newblock}{\relax}
\providecommand{\bibinfo}[2]{#2}
\providecommand{\BIBentrySTDinterwordspacing}{\spaceskip=0pt\relax}
\providecommand{\BIBentryALTinterwordstretchfactor}{4}
\providecommand{\BIBentryALTinterwordspacing}{\spaceskip=\fontdimen2\font plus
\BIBentryALTinterwordstretchfactor\fontdimen3\font minus
  \fontdimen4\font\relax}
\providecommand{\BIBforeignlanguage}[2]{{%
\expandafter\ifx\csname l@#1\endcsname\relax
\typeout{** WARNING: IEEEtran.bst: No hyphenation pattern has been}%
\typeout{** loaded for the language `#1'. Using the pattern for}%
\typeout{** the default language instead.}%
\else
\language=\csname l@#1\endcsname
\fi
#2}}
\providecommand{\BIBdecl}{\relax}
\BIBdecl

\bibitem{sigona2018voice}
F.~Sigona, ``Voice biometrics technologies and applications for healthcare: an
  overview.'' \emph{JDREAM. Journal of interDisciplinary REsearch Applied to
  Medicine}, vol.~2, no.~1, pp. 5--16, 2018.

\bibitem{singh2012applications}
N.~Singh, R.~Khan, and R.~Shree, ``Applications of speaker recognition,''
  \emph{Procedia engineering}, vol.~38, pp. 3122--3126, 2012.

\bibitem{changmy}
Y.-T. Chang and M.~J. Dupuis, ``My voiceprint is my authenticator: A two-layer
  authentication approach using voiceprint for voice assistants,'' in
  \emph{2019 IEEE SmartWorld, Ubiquitous Intelligence \& Computing, Advanced \&
  Trusted Computing, Scalable Computing \& Communications, Cloud \& Big Data
  Computing, Internet of People and Smart City Innovation
  (SmartWorld/SCALCOM/UIC/ATC/CBDCom/IOP/SCI)}.\hskip 1em plus 0.5em minus
  0.4em\relax IEEE, 2019, pp. 1318--1325.

\bibitem{sdsvc2020plan}
H.~Zeinali, K.~A. Lee, J.~Alam, and L.~Burget, ``Short-duration speaker
  verification (sdsv) challenge 2020: the challenge evaluation plan.'' arXiv
  preprint arXiv:1912.06311, Tech. Rep., 2020.

\bibitem{dehak2010front}
N.~Dehak, P.~J. Kenny, R.~Dehak, P.~Dumouchel, and P.~Ouellet, ``Front-end
  factor analysis for speaker verification,'' \emph{IEEE Transactions on Audio,
  Speech, and Language Processing}, vol.~19, no.~4, pp. 788--798, 2010.

\bibitem{kenny2013plda}
P.~Kenny, T.~Stafylakis, P.~Ouellet, M.~J. Alam, and P.~Dumouchel, ``Plda for
  speaker verification with utterances of arbitrary duration,'' in \emph{2013
  IEEE International Conference on Acoustics, Speech and Signal
  Processing}.\hskip 1em plus 0.5em minus 0.4em\relax IEEE, 2013, pp.
  7649--7653.

\bibitem{larcher2013phonetically}
A.~Larcher, K.~A. Lee, B.~Ma, and H.~Li, ``Phonetically-constrained plda
  modeling for text-dependent speaker verification with multiple short
  utterances,'' in \emph{2013 IEEE International Conference on Acoustics,
  Speech and Signal Processing}.\hskip 1em plus 0.5em minus 0.4em\relax IEEE,
  2013, pp. 7673--7677.

\bibitem{stafylakis2013text}
T.~Stafylakis, P.~Kenny, P.~Ouellet, J.~Perez, M.~Kockmann, and P.~Dumouchel,
  ``Text-dependent speaker recognition using plda with uncertainty
  propagation,'' \emph{matrix}, vol. 500, no.~1, 2013.

\bibitem{zeinali2016vector}
H.~Zeinali, H.~Sameti, L.~Burget, J.~Cernock{\`y}, N.~Maghsoodi, and
  P.~Matejka, ``i-vector/hmm based text-dependent speaker verification system
  for reddots challenge.'' in \emph{InterSpeech}, 2016, pp. 440--444.

\bibitem{zeinali2017hmm}
H.~Zeinali, H.~Sameti, and L.~Burget, ``Hmm-based phrase-independent i-vector
  extractor for text-dependent speaker verification,'' \emph{IEEE/ACM
  Transactions on Audio, Speech, and Language Processing}, vol.~25, no.~7, pp.
  1421--1435, 2017.

\bibitem{variani2014deep}
E.~Variani, X.~Lei, E.~McDermott, I.~L. Moreno, and J.~Gonzalez-Dominguez,
  ``Deep neural networks for small footprint text-dependent speaker
  verification,'' in \emph{2014 IEEE International Conference on Acoustics,
  Speech and Signal Processing (ICASSP)}.\hskip 1em plus 0.5em minus
  0.4em\relax IEEE, 2014, pp. 4052--4056.

\bibitem{lei2014novel}
Y.~Lei, N.~Scheffer, L.~Ferrer, and M.~McLaren, ``A novel scheme for speaker
  recognition using a phonetically-aware deep neural network,'' in \emph{2014
  IEEE International Conference on Acoustics, Speech and Signal Processing
  (ICASSP)}.\hskip 1em plus 0.5em minus 0.4em\relax IEEE, 2014, pp. 1695--1699.

\bibitem{zeinali2016deep}
H.~Zeinali, L.~Burget, H.~Sameti, O.~Glembek, and O.~Plchot, ``Deep neural
  networks and hidden markov models in i-vector-based text-dependent speaker
  verification.'' in \emph{Odyssey}, 2016, pp. 24--30.

\bibitem{guo2017cnn}
J.~Guo, U.~A. Nookala, and A.~Alwan, ``Cnn-based joint mapping of short and
  long utterance i-vectors for speaker verification using short utterances.''
  in \emph{INTERSPEECH}, 2017, pp. 3712--3716.

\bibitem{guo2018deep}
J.~Guo, N.~Xu, K.~Qian, Y.~Shi, K.~Xu, Y.~Wu, and A.~Alwan, ``Deep neural
  network based i-vector mapping for speaker verification using short
  utterances,'' \emph{Speech Communication}, vol. 105, pp. 92--102, 2018.

\bibitem{heigold2016end}
G.~Heigold, I.~Moreno, S.~Bengio, and N.~Shazeer, ``End-to-end text-dependent
  speaker verification,'' in \emph{2016 IEEE International Conference on
  Acoustics, Speech and Signal Processing (ICASSP)}.\hskip 1em plus 0.5em minus
  0.4em\relax IEEE, 2016, pp. 5115--5119.

\bibitem{zhang2016end}
S.-X. Zhang, Z.~Chen, Y.~Zhao, J.~Li, and Y.~Gong, ``End-to-end attention based
  text-dependent speaker verification,'' in \emph{2016 IEEE Spoken Language
  Technology Workshop (SLT)}.\hskip 1em plus 0.5em minus 0.4em\relax IEEE,
  2016, pp. 171--178.

\bibitem{wan2018generalized}
L.~Wan, Q.~Wang, A.~Papir, and I.~L. Moreno, ``Generalized end-to-end loss for
  speaker verification,'' in \emph{2018 IEEE International Conference on
  Acoustics, Speech and Signal Processing (ICASSP)}.\hskip 1em plus 0.5em minus
  0.4em\relax IEEE, 2018, pp. 4879--4883.

\bibitem{devlin2018bert}
J.~Devlin, M.-W. Chang, K.~Lee, and K.~Toutanova, ``Bert: Pre-training of deep
  bidirectional transformers for language understanding,'' \emph{arXiv preprint
  arXiv:1810.04805}, 2018.

\bibitem{schneider2019wav2vec}
S.~Schneider, A.~Baevski, R.~Collobert, and M.~Auli, ``wav2vec: Unsupervised
  pre-training for speech recognition,'' \emph{arXiv preprint
  arXiv:1904.05862}, 2019.

\bibitem{chung2019unsupervised}
Y.-A. Chung, W.-N. Hsu, H.~Tang, and J.~Glass, ``An unsupervised autoregressive
  model for speech representation learning,'' \emph{arXiv preprint
  arXiv:1904.03240}, 2019.

\bibitem{chung2020generative}
Y.-A. Chung and J.~Glass, ``Generative pre-training for speech with
  autoregressive predictive coding,'' in \emph{ICASSP}, 2020.

\bibitem{oord2018representation}
A.~v.~d. Oord, Y.~Li, and O.~Vinyals, ``Representation learning with
  contrastive predictive coding,'' \emph{arXiv preprint arXiv:1807.03748},
  2018.

\bibitem{liu2015deep}
Y.~Liu, Y.~Qian, N.~Chen, T.~Fu, Y.~Zhang, and K.~Yu, ``Deep feature for
  text-dependent speaker verification,'' \emph{Speech Communication}, vol.~73,
  pp. 1--13, 2015.

\bibitem{sarkar2019time}
A.~K. Sarkar, Z.-H. Tan, H.~Tang, S.~Shon, and J.~Glass, ``Time-contrastive
  learning based deep bottleneck features for text-dependent speaker
  verification,'' \emph{Ieee/acm Transactions on Audio, Speech, and Language
  Processing}, vol.~27, no.~8, pp. 1267--1279, 2019.

\bibitem{o1988linear}
D.~O'Shaughnessy, ``Linear predictive coding,'' \emph{IEEE potentials}, vol.~7,
  no.~1, pp. 29--32, 1988.

\bibitem{graves2006connectionist}
A.~Graves, S.~Fern{\'a}ndez, F.~Gomez, and J.~Schmidhuber, ``Connectionist
  temporal classification: labelling unsegmented sequence data with recurrent
  neural networks,'' in \emph{Proceedings of the 23rd international conference
  on Machine learning}, 2006, pp. 369--376.

\bibitem{Nagrani17}
A.~Nagrani, J.~S. Chung, and A.~Zisserman, ``Voxceleb: a large-scale speaker
  identification dataset,'' in \emph{INTERSPEECH}, 2017.

\bibitem{zeinali2018deepmine}
H.~Zeinali, H.~Sameti, and T.~Stafylakis, ``Deepmine speech processing
  database: Text-dependent and independent speaker verification and speech
  recognition in persian and english.'' in \emph{Odyssey}, 2018, pp. 386--392.

\bibitem{zeinali2019multi}
H.~Zeinali, L.~Burget, J.~{\v{C}}ernock{\`y} \emph{et~al.}, ``A multi purpose
  and large scale speech corpus in persian and english for speaker and speech
  recognition: the deepmine database,'' \emph{arXiv preprint arXiv:1912.03627},
  2019.

\bibitem{snyder2018x}
D.~Snyder, D.~Garcia-Romero, G.~Sell, D.~Povey, and S.~Khudanpur, ``X-vectors:
  Robust dnn embeddings for speaker recognition,'' in \emph{2018 IEEE
  International Conference on Acoustics, Speech and Signal Processing
  (ICASSP)}.\hskip 1em plus 0.5em minus 0.4em\relax IEEE, 2018, pp. 5329--5333.

\bibitem{snyder2015musan}
D.~Snyder, G.~Chen, and D.~Povey, ``Musan: A music, speech, and noise corpus,''
  \emph{arXiv preprint arXiv:1510.08484}, 2015.

\bibitem{povey2011kaldi}
D.~Povey, A.~Ghoshal, G.~Boulianne, L.~Burget, O.~Glembek, N.~Goel,
  M.~Hannemann, P.~Motlicek, Y.~Qian, P.~Schwarz \emph{et~al.}, ``The kaldi
  speech recognition toolkit,'' in \emph{IEEE 2011 workshop on automatic speech
  recognition and understanding}, no. CONF.\hskip 1em plus 0.5em minus
  0.4em\relax IEEE Signal Processing Society, 2011.

\bibitem{martin2010nist}
A.~F. Martin and C.~S. Greenberg, ``The nist 2010 speaker recognition
  evaluation,'' in \emph{Eleventh Annual Conference of the International Speech
  Communication Association}, 2010.

\end{thebibliography}
\end{document}